\documentclass[%
 reprint,
nofootinbib,
 amsmath,amssymb,
 aps,
]{revtex4-2}

\usepackage{graphicx}
\usepackage{dcolumn}
\usepackage{bm}
\usepackage{bbold}
\usepackage{multirow}
\usepackage{caption}
\usepackage{bbding}
\usepackage{pifont}
\usepackage{wasysym}
\usepackage{amssymb}

\usepackage{xcolor}
\usepackage{tikz-feynman} 

\usepackage{graphicx}
\usepackage{dcolumn}
\usepackage{bm}
\usepackage{amsmath}
\usepackage{amssymb}
\usepackage{bbm}
\usepackage{multirow}
\usepackage{comment}
\usepackage{caption}

\usepackage[T1]{fontenc} 
\newcommand{\ie}{i.e.,}
\newcommand{\gp}{g_{Y}}

\newcommand{\gzpp}{g_{Z^{\prime\prime}}}
\newcommand{\zp}{Z^{\prime}}
\newcommand{\zpp}{Z^{\prime\prime}}

\newcommand{\ha}{\hat{\alpha}_R}
\newcommand{\hb}{\frac{g_{L}}{g_X}}
\newcommand{\hbLR}{\frac{g_{L}}{g_X}}
 \newcommand{\cosx}{\cos\theta}
 \newcommand{\sinx}{\sin\theta}

 \newcommand{\cosy}{\cos\phi}
 \newcommand{\siny}{\sin\phi}
 
 \newcommand{\cosz}{\cos\omega}
 \newcommand{\sinz}{\sin\omega}
 
\usepackage{caption}
\usepackage{capt-of} 

\begin{document}

\preprint{APS/123-QED}

\title{Non-Universal Flipped Trinification Models with Arbitrary 
$\beta$}

\author{Richard H. Benavides}
\email{richardbenavides@itm.edu.co}
\affiliation{ Instituto Tecnol\'{o}gico Metropolitano, Facultad de Ciencias Exactas y Aplicadas,  \textit{\small Calle 73 N° 76-354 vía el volador, Medell\'{i}n, Colombia.}}
\author{Yithsbey Giraldo}
\email{yithsbey@gmail.com}
\affiliation{Departamento de Física, Universidad de Nariño, \textit{\small Calle 18 Carrera 50, A.A. 1175, San Juan de Pasto, Colombia}}

\author{Eduardo Rojas}
\email{eduro4000@gmail.com}
\affiliation{Departamento de Física, Universidad de Nariño, \textit{\small Calle 18 Carrera 50, A.A. 1175, San Juan de Pasto, Colombia}}

\date{\today}

\begin{abstract}
We explore the recently proposed gauge symmetry
\( SU(3)_C \otimes SU(3)_L \otimes SU(3)_R \otimes U(1)_X \), 
which naturally embeds both the Left-Right symmetric model and the 3-3-1 model as subgroups. Within this unified framework, we propose four families of leptons and quarks. A detailed analysis of their contributions to gauge anomaly cancellation is carried out for a general value of the parameter $\beta$.

From this analysis, eight non-universal anomaly-free three-family models  and four non-universal two-family anomaly free sets were identified. The three-family models offer realistic extensions of the Standard Model, retaining several appealing features of the 3-3-1 models, while the two-family sets provide flexibility for constructing models with even numbers of families. 
We also report LHC bounds on the $Z'$ mass for the particular case $\beta = -1/\sqrt{3}$, considering all possible combinations of lepton and quark families. These limits exhibit a strong dependence on the mixing parameter $\theta$, which enters the couplings of Standard Model fermions to the $Z'$ boson.
\end{abstract}

\maketitle


\section{\label{sec:intro}Introduction}
The Standard Model (SM) does not fully explain why electric charge is quantized, that is, why all observed charges are integer multiples of $e/3$. Anomaly cancelation imposes certain restrictions on hypercharges, but these are not sufficient to uniquely determine the SM charges, even adding right-handed neutrinos~\cite{Foot:1992ui,Nowakowski:1992ff}. 
So, the standard lore says that it may be necessary to embed the SM into a larger symmetry group to address this issue~\cite{Pati:1974yy,Georgi:1974sy}.

An extension that has been studied recently in the literature is flipped-trinification~(FT)~\cite{Reig:2016tuk, Hati:2017aez, Dong:2017zxo,Huong:2018ytz,Dinh:2019jdg}, which is based on the  $SU(3)_C\otimes SU(3)_L\otimes SU(3)_R\otimes U(1)_X$~(3-3-3-1) gauge group; this class of models can explain the number of families of the Standard Model as 3-3-1 models do~\cite{Singer:1980sw,Pisano:1992bxx,Frampton:1992wt,Benavides:2021pqx,Suarez:2023ozu}, and the origin of parity violation as it happens in the Left-Right models~(LR)~\cite{Pati:1974yy,Mohapatra:1974hk}. This class of models is also suitable for explaining dark-matter stability.

The left-right symmetric models are among the simplest extensions of the SM. This class of models represents an ideal theoretical framework for understanding the origin of parity violation in gauge theories, neutrino masses, and dark matter. Similarly, 3-3-1 models have been used for many years to explain neutrino masses and dark matter.  Additionally, they can provide a relation between the number of families and the number of colors~\cite{Frampton:1992wt}. FT manages to integrate in a single model the advantages of the LR and the 3-3-1 models, creating a theoretical framework that includes two of the most motivated extensions of the SM.

By a suitable choice of the Higgs sector, it is possible to break the trinification gauge group to $SU(3)_C\otimes SU(2)_L\otimes U(1)_{L8+R8}\otimes U(1)_R \otimes U(1)_X$~\footnote{Where the generator of $U(1)_{L8+R8}$ in dimension 3 is given by $\frac{1}{2}\left(\lambda^8(f_L)+\lambda^8(f_R)\right)$, with $\lambda^a(f)$ is the representation of the $SU(3)$ group in the fermions $f$.} 
We provide a classification of the viable models that can be derived from this symmetry, subject to certain assumptions.\\

In Section~\ref{sec:QLfamilies}, we construct all possible families for models based on this symmetry and calculate their contribution to the anomalies. From these results, we obtain several non-universal 3-3-3-1 models.
In Section~\ref{sec:zpcharges} we calculate the $Z'$ electroweak charges. 
In Section~\ref{sec:constraints}, we compute and summarize the collider constraints for all combinations of quark and lepton families considered in our analysis.
\section{Quark and Lepton families\label{sec:QLfamilies}}
In this section, we construct sets of fermions that include at least one SM quark or lepton family. These sets also contain exotic particles, as will be discussed below.
Each family must have at least one quark and one lepton triplet to accommodate a left quark doublet and a lepton doublet from the Standard Model. It is also necessary to have $SU(2)_L$ singlets to accommodate the right-handed components of the charged particles. In $SU(3)_L$ for each $3_L$ triplet, there exists the conjugate representation $3^*_L$, and in both triplets it is possible to accommodate the left doublets of the Standard Model. Due to $SU(3)_R$, there also exist right $3_R$ triplets and their respective conjugate representations $3_R^*$, so that it is possible to put the right-handed singlets of the Standard Model in these triplets. The charge operator is given by
\begin{align}
Q=T_{3L}+Y=T_{3L}+T_{3R}+\beta\left(T_{8L}+T_{8R}\right)+X\mathbf{1}\ , \label{eq:Qoperator}   
\end{align}
where $T_{aL,R}=\frac{1}{2}\lambda^{a}_{L,R}$ with 
 $\lambda^{a}_{L,R}=\lambda^{a}P_{L,R}$, such that the $\lambda^a$'s 
are the Gell-Mann matrices.
By treating the charge \( q \) of the third component of the lepton triplet as a free parameter, we determine the values of \( \beta \) and \( X \) from the relation \( Q = \text{diag}(0, -1, q) \), such that $X=\frac{q-1}{3}\ $ and 
 \begin{align*}
\beta = -\frac{1+2q}{\sqrt{3}}\ .
 \end{align*}
 It is important to use the same value of $\beta$ for all fermion representations to maintain a unique adjoint representation.  In our case, this is not an issue, since the electric charge is given by $ q = -\frac{1 + \sqrt{3}\beta}{2}$.
 In similar way we can obtain the charges of the 
 conjugated representations from $Q^{\text{conj}}_{L,R}=-\frac{1}{2}\lambda^{3}_{L,R}-\frac{\beta}{2}\lambda^{8}_{L,R}+X_{L,R}\mathbf{1}$.
 For the lepton triplet, with $X=-\frac{q+2}{3}$, we have 
 $Q^{\text{conj}}_{L,R}=\text{diag}(-1,0,-1-q)$.
 Similar results hold for the quark triplet with $X=\frac{q+1}{3}$,  $Q_{L,R}=\text{diag}(2/3,-1/3,q+\frac{2}{3}) $  and $Q^{\text{conj}}_{L,R}=\text{diag}(-1/3,+2/3,-\frac{1}{3}-q)$ 
 with $X= -\frac{q}{3}$. 
 
From these considerations, we derive four viable families of \( S_{Qi} \) quarks and four of \( S_{Li} \) leptons, which are listed below.

\subsection*{Families in the Lepton sector }
The first generation of leptons is assigned to the $\mathbf{3}_L$ and $\mathbf{3}_R$ representations. Owing to the imposed left-right (L-R) symmetry, the $U(1)_X$ charge is identical for both representations.
\begin{align*}
    S_{L1} &= \begin{pmatrix}
        \nu_L \\
        e_L \\
        E_L^q
    \end{pmatrix}\ , 
    \begin{pmatrix}
        \nu_R \\
        e_R \\
        E_R^q
    \end{pmatrix}.  
 \end{align*}
In $S_{L1}= \mathbf{3}_L\cup\mathbf{3}_R$, the left-handed SM lepton doublets are embedded in $ (1, 3, 1, \frac{q-1}{3})$ and the right-handed SM lepton components in $ (1,1,3, \frac{q-1}{3})$.
 \begin{align*}
       S_{L2}& = \begin{pmatrix}
        \nu_L \\
        e_L \\
        E_L^q
    \end{pmatrix}, 
    \begin{pmatrix}
        e_R \\
        \nu_R \\
        E_R^{-q-1}
    \end{pmatrix}, E^q_R, E^{-q-1}_L.
 \end{align*}
For $S_{L2} = \mathbf{3}_L \cup \mathbf{3}_R^* \cup \mathbf{1}_R \cup \mathbf{1}_L$, the SM left-handed lepton doublets are embedded in  
$(1, 3, 1, \frac{q - 1}{3})$,  
the right-handed leptons are in $(1, 1, 3^*, -\frac{q + 2}{3})$,  
and the chirality flipped  components of the third element of the triplets  in  
$(1, 1, 1, q)$ and $(1, 1, 1, -q - 1)$, respectively.
\\
For \( q = -\frac{1}{2} \)~(which is equivalent to $\beta=0$), the third components of the left- and right-handed triplets have the same electric charge, eliminating the need to introduce singlets.
  \begin{align*}
    S_{L3} &= \begin{pmatrix}
        e_L \\
        \nu_L \\
        E_L^{-q-1}
    \end{pmatrix}, \quad
    \begin{pmatrix}
        \nu_R \\
        e_R \\
        E_R^{q}
    \end{pmatrix}, \quad E^{-q-1}_R, \quad E^{q}_L\ . 
 \end{align*}
 For $S_{L3} = \mathbf{3}_L^* \cup \mathbf{3}_R \cup \mathbf{1}_R \cup \mathbf{1}_L$, the SM left-handed lepton doublets are embedded in $\quad(1, 3^*, 1, -\frac{q+2}{3})$, 
 the right-handed lepton in 
 $(1, 1, 3, \frac{q-1}{3})$  
 and the chirality flipped  components of the third element of the triplets  in  $(1, 1, 1, -q-1)$ and $(1, 1, 1, q)$ with $q\neq -\frac{1}{2}$. 
 \begin{align*}
    S_{L4} &= \begin{pmatrix}
        e_L \\
        \nu_L \\
        E_L^{-q-1}
    \end{pmatrix}\ , 
     \begin{pmatrix}
        e_R \\
        \nu_R \\
        E_R^{-q-1}
    \end{pmatrix}\ .
  \end{align*}
For $S_{L4} = \mathbf{3}_L^* \cup \mathbf{3}_R^*$, the SM left-handed lepton doublets are embedded in
$(1, 3^*, 1, -\frac{q+2}{3})$
and the right-handed leptons in
$(1,1, 3^*, -\frac{q+2}{3})$.
\subsection*{Families in the quark sector}
Proceeding in the same way as in the case of leptons, we have:
{\small
\begin{align*}
    S_{Q1} &= \begin{pmatrix}
        u_L \\
        d_L \\
        Q_L^{q+2/3}
    \end{pmatrix}, \quad \begin{pmatrix}
        u_R \\
        d_R \\
        Q_R^{q+2/3}
    \end{pmatrix}\ .
\end{align*}
For $S_{Q1} = \mathbf{3}_L \cup \mathbf{3}_R$, the SM left-handed quark doublets are embedded in
$(3, 3, 1, \frac{q + 1}{3})$ and the right-handed quarks in $(3, 1, 3, \frac{q + 1}{3})$\ .
\begin{align*}
    S_{Q2} &= \begin{pmatrix}
        u_L \\
        d_L \\
        Q_L^{q+2/3}
   \end{pmatrix},  \begin{pmatrix}
        d_R \\
        u_R \\
        Q_R^{-q-1/3}
    \end{pmatrix},  Q_R^{q+2/3}, Q_L^{-q-1/3}\ .
\end{align*}
For $S_{Q2} = \mathbf{3}_L \cup \mathbf{3}_R^*\cup\mathbf{1}_R\cup \mathbf{1}_L$, the SM left-handed quark doublets are embedded in $(3, 3, 1, \frac{q+1}{3})$,
the right-handed quarks in $(3, 1,3^*, -\frac{q}{3})$ and the chirally flipped third components of the quark triplets in 
$(3, 1,1, q+2/3)$, $(3, 1,1, -q-1/3)$, with $q\neq -\frac{1}{2}$\ .
\begin{align*}
        S_{Q3}&= \begin{pmatrix}
        d_L \\
        u_L \\
        Q_L^{-q-1/3}
    \end{pmatrix}, \begin{pmatrix}
        u_R \\
        d_R \\
        Q_R^{q+2/3}
    \end{pmatrix},  Q_R^{-q-1/3}, Q_L^{q+2/3}\ .
\end{align*}
For $S_{Q3} = \mathbf{3}_L^* \cup \mathbf{3}_R\cup\mathbf{1}_R\cup \mathbf{1}_L$, the SM left-handed quark doublets are embedded in $(3, 3^*, 1, -\frac{q}{3})$,
the right-handed quarks in  $(3, 1,3, \frac{q+1}{3})$  and the chirally flipped third components of the quark triplets in 
$(3, 1,1,-q-1/3 )$ and $(3, 1,1, q+2/3)$, with $q\neq -\frac{1}{2}$\ .
\begin{align*}
      S_{Q4}& = \begin{pmatrix}
        d_L \\
        u_L \\
        Q_L^{-q-1/3}
    \end{pmatrix}, \quad \begin{pmatrix}
        d_R \\
       u_R \\
        Q_R^{-q-1/3}
    \end{pmatrix}\ .
    \end{align*}}
For $S_{Q4} = \mathbf{3}_L^* \cup \mathbf{3}_R^*$, the SM left-handed quark doublets are embedded in $(3, 3^*,1, -\frac{q}{3})$,
the right-handed quarks in  $(3, 1,3^*, -\frac{q}{3})$\ . 
\begin{widetext}
 {\footnotesize
\scalebox{0.9}{
\begin{tabular}{|m{2cm}|m{.1cm}m{.8cm}m{0.8cm}m{.1cm}m{1.2cm}m{.8cm}m{1cm}m{.1cm}|}\hline
Anomalies & $S_{L1}$ & $S_{L2}$ &  \hspace{-4mm}$S_{L3}$ & $S_{L4}$ & $S_{Q1}$ & \hspace{-4mm}$S_{Q2}$ & $S_{Q3}$& $S_{Q4}$ \\ \hline
$[SU(3)_C]^2\otimes~U(1)_X$ &  0 & 0 &  0 &  0 &  0 & \hspace{-4mm} 0 &  0 &0\\
$[SU(3)_L]^2\otimes~U(1)_X$ & $\frac{q-1}{3}$ &\hspace{-3mm} $\frac{q-1}{3}$ &  \hspace{-6mm}$-\frac{(q+2)}{3}$ &$-\frac{(q+2)}{3}$ & $1+q$ &\hspace{-4mm} $1+q$ & $-q$&$-q$\\
$[SU(3)_R]^2\otimes~U(1)_X$ & $\frac{q-1}{3}$ &\hspace{-6mm} $-\frac{(q+2)}{3}$ &   \hspace{-4mm}$\frac{q-1}{3}$ &$-\frac{(q+2)}{3}$ & $1+q$ & \hspace{-4mm}$-q$  &$1+q$&$-q$\\
$[\text{Grav}]^2\otimes~U(1)_X$    & 0 & 0 & 0 & 0 & 0 &\hspace{-4mm}0 & 0&0\\
$[U(1)_X]^3$        & 0 &\hspace{-8mm} $-\frac{2}{9}(1+2q)^3$ & \hspace{-6mm}$\frac{2}{9}(1+2q)^3$  &  $0$ & 0 &  \hspace{-8mm}$-\frac{2}{3}(1+2q)^3$ & \hspace{-4mm}$\frac{2}{3}(1+2q)^3$&0\\
 $[SU(3)_L]^3$ &&&&&&&&\\[-5mm]
 and & 2 & $0$ & $0$ & $-2$ & 6 & $0$ & 0&  $-6$\\[-5mm]
 $[SU(3)_R]^3$ &&&&&&&& \\
\hline
\end{tabular}}
\captionof{table}{\small
Contribution to the anomalies from each quark family ($S_{Qi}$) and lepton family ($S_{Li}$) in 3-3-3-1 models.}  
\label{tab:anomalias}
} 
\end{widetext}

\section{Irreducible anomaly-free sets \label{sec:afs}}

By combining the families of quarks \( S_{Qi} \) and leptons \( S_{Li} \), according to Table~\ref{tab:anomalias}, it is possible to construct eight three-generation models that produce anomaly-free particle spectra. The corresponding eight three-family models are: \\ 

$M_1: 3 S_{L4}+S_{Q1}+S_{Q2}+S_{Q3}$ \ ,\\

$M_2: 3 S_{L4}+2 S_{Q1}+S_{Q4}$\ ,\\

$M_3: 3 S_{L3}+2 S_{Q2}+S_{Q3}$\ ,\\

$M_4: 3 S_{L3}+S_{Q1}+S_{Q2}+S_{Q4}$\ ,\\

$M_5: 3 S_{L2}+S_{Q2}+2 S_{Q3}$\ ,\\

$M_6: 3 S_{L2}+S_{Q1}+S_{Q3}+S_{Q4}$\ ,\\

$M_7: 3 S_{L1}+S_{Q2}+S_{Q3}+S_{Q4}$\ ,\\

$M_8: 3 S_{L1}+S_{Q1}+2 S_{Q4}$\ .\\

It is also possible to construct anomaly-free sets containing two generations, which may serve as building blocks for models with an even number of generations, such as four, six, and so on. However, since these models do not account for the observed three families in the SM, they have only been sparsely explored in the literature. However, it is important to note that models that incorporate additional generations can play a significant role in various phenomenological contexts.
The irreducible anomaly-free sets  of two families are: \\

${S_{L2}}+{S_{L3}}+{S_{Q2}}+{S_{Q3}}$\ ,\\

${S_{L2}}+{S_{L3}}+{S_{Q1}}+{S_{Q4}}$\ ,\\

${S_{L1}}+{S_{L4}}+{S_{Q2}}+{S_{Q3}}$\ ,\\

${S_{L1}}+{S_{L4}}+{S_{Q1}}+{S_{Q4}}$\ .\\

By introducing two such sets, a four-family model emerges. Three of these families align with those of the Standard Model, while the fourth is considered exotic. There exists extensive literature on top-prime models~(see~\cite{Franceschini:2023nlp} and references therein), encompassing both theoretical frameworks and phenomenological analyses.

\subsection{Embeddings}

As shown by the previous results, models based on irreducible sets of three families exhibit universality in the lepton sector. In contrast, in the quark sector, universality holds for at most two families. Therefore, we restrict our analysis to embeddings in the quark sector, identifying the two universal families with the first and second generations of the Standard Model.

\begin{widetext}
\begin{center}
\scalebox{0.84}{
\begin{tabular}{|l|ccccc|}\hline
Model&Lepton content&SM Quark Embeddings &$2+1$&FCNC&LHC-Lower limit (TeV)
\\
&&&&&\\[-6.5mm]
\hline
&&&&&\\[-3.5mm]
\multirow{2}{*}{$M_1$}&$3 S_{L4}$&  $S_{Q_1}+S_{Q_2}+S_{Q_3}$&$\times$&
\checkmark& ED\\
&&&&&\\[-3.5mm]\hline
&&&&&\\[-3.5mm]
\multirow{1}{*}{$M_2$}&$3 S_{L4}$&  $2S_{Q_1}^{ud,cs}+S_{Q_4}^{tb}$&\checkmark&$\times$
&4.3\\
&&&&&\\[-3.5mm]\hline
&&&&&\\[-3.5mm]
\multirow{1}{*}{$M_3$}&$3 S_{L3}$&  $2S_{Q_2}^{ud,cs}+S_{Q_3}^{tb}$&\checkmark&$\times$
&4\\
&&&&&\\[-3.5mm]\hline
&&&&&\\[-3.5mm]
\multirow{1}{*}{$M_4$}&$3 S_{L3}$&   $S_{Q_1}+S_{Q_2}+S_{Q_4}$&$\times$&\checkmark
&ED\\
&&&&&\\[-3.5mm]\hline
&&&&&\\[-3.5mm]
\multirow{1}{*}{$M_5$}&$3 S_{L2}$&   $2S_{Q_3}^{ud,cs}+S_{Q_2}^{tb}$&\checkmark&$\times$
&4.5\\
&&&&&\\[-3.5mm]\hline
&&&&&\\[-3.5mm]
\multirow{1}{*}{$M_6$}&$3 S_{L2}$& $S_{Q_1}+S_{Q_3}+S_{Q_4}$&$\times$&\checkmark
&ED\\
&&&&&\\[-3.5mm]\hline
&&&&&\\[-3.5mm]
\multirow{1}{*}{$M_7$}&$3 S_{L1}$&$S_{Q_2}+S_{Q_3}+S_{Q_4}$&$\times$&\checkmark
&ED\\
&&&&&\\[-3.5mm]\hline
&&&&&\\[-3.5mm]
\multirow{1}{*}{$M_8$}&$3 S_{L1}$&   $2S_{Q_4}^{ud,cs}+S_{Q_1}^{tb}$&\checkmark&$\times$
&4.3\\[3.5mm]
\hline
\end{tabular}
}
\captionof{table}{\small
Embeddings of the Standard Model quarks within each family are shown.
The superscript denotes the quark content of each family.
The lepton sector is universal and does not require an explicit embedding. A check mark indicates that a configuration with at least two families (2+1) sharing identical $Z^\prime$ charges is possible. LHC constraints are reported only for embeddings where identical $Z^\prime$ charges can be assigned to the first two families. In models with three non-universal quark families, Flavor-Changing Neutral Currents~(FCNC) arise between the first two generations, in conflict with experimental constraints. In all cases, the models exhibit universality in the lepton sector.  ED stands for Embedding Dependent.
In these cases, the lower limit depends on which families are identified with the first generation of the SM.
}  
\label{tabl5}
\end{center}
\end{widetext}

\section{Electroweak $Z'$ charges\label{sec:zpcharges}}

From the charge operator~\eqref{eq:Qoperator} 
we can obtain the corresponding  low-energy gauge group $G=SU(3)_C\otimes SU(2)_L\otimes SU(2)_R\otimes U(1)_{8L+8R}\otimes U(1)_X$.  
The neutral current Lagrangian for these models is:
\begin{align}\label{eq:LRLagrangian}
-\mathcal{L}_{NC}
=&g_L J_{3L\mu}A_{3L}^{\mu}+g_{L}\left(J_{8L\mu}+J_{8R\mu}\right)A_{8}^{\mu}\notag\\
+&g_{X} J_{X\mu}A_{X}^{\mu}
+g_R J_{3R\mu}A_{3R}^{\mu}\ ,\notag\\
=&g_L J_{3L\mu}A_{3L}^{\mu}+g^{\prime}J_{Y\mu} B^\mu\notag\\
+&g_{Z'}J_{Z'\mu} Z^{\prime\mu}
 +g_{Z''}J_{Z''\mu} Z^{\prime\prime\mu}\ .
\end{align}
By means of an orthogonal matrix we can rotate from the left-right basis of the Neutral Current~(NC) vector bosons 
to the $(B,Z')$ basis \ie
\begin{equation}\label{rotation1}
\begin{pmatrix}
A^\mu_{3L} \\
A^\mu_{8} \\
A^\mu_{X}\\ 
A^\mu_{3R} \\
\end{pmatrix} =
\mathcal{O} 
\begin{pmatrix}
A^\mu_{3L} \\
B^\mu \\
Z^{\prime\mu} \\
Z^{\prime\prime\mu} 
\end{pmatrix}
=
\begin{pmatrix}
1 &0 &0 &0 \\
0 &O_{11}  &O_{12}  &O_{13}  \\
0 &O_{21}  &O_{22}  &O_{23}  \\
0 &O_{31}  &O_{32}  &O_{33}  
\end{pmatrix}
\begin{pmatrix}
A^\mu_{3L} \\
B^\mu \\
Z^{\prime\mu} \\
Z^{\prime\prime\mu} 
\end{pmatrix}\ ,
\end{equation} 
In order to keep invariant the Lagrangian, the currents must transform with the same orthogonal matrix 
\begin{align}
\gp J^\mu_Y                
&= g_{L} J^\mu_{8}O_{11} + g_{X} J^\mu_{X} O_{21} + g_R J^\mu_R O_{31}\ , \label{Eq1}\\ 
g_{\zp} J_{\zp}^\mu    
&= g_{L} J^\mu_{8}O_{12} + g_{X} J^\mu_{X} O_{22} + g_R J^\mu_R O_{32}\ , \label{Eq1b}\\
g_{\zpp}J_{\zpp}^\mu   
&= g_{L} J^\mu_{8}O_{13} + g_{X} J^\mu_{X} O_{23} + g_R J^\mu_R O_{33}\ . \label{Eq1c} 
\end{align}
From the charge operator, we have 
\begin{align}
Y= \beta\left(T_{8L}+T_{8R}\right)+T_{3R}+X\mathbf{1}\ .    
\end{align}
It is also well known that 
$Y=T_{3R}+\frac{1}{2}(B-L)$, such that 
$\frac{1}{2}(B-L)=\beta\left(T_{8L}+T_{8R}\right)+X\mathbf{1} $.
Which implies the relation
\begin{align}
J_Y= \beta\left(J_{8L}+J_{8R}\right)+J_{3R}+J_X\ .    
\end{align}
By comparing these expressions we  obtain 
\begin{align}\label{eq:angles1}
\frac{g_{L}}{g_Y}O_{11}= \beta\ ,
\hspace{0.5cm}
\frac{g_X}{g_Y}O_{21}= 1\ ,
\hspace{0.5cm}
\frac{g_{R}}{g_Y}O_{31}= 1\ .
\end{align}
As is well known, an orthogonal matrix satisfies
$O_{ij}O^{T}_{jk}=\delta_{ik}$. For $i=k=1$ we have   
$O_{1j}O^{T}_{j1}=O_{1j}O_{1j}=O_{11}^2+O_{12}^2+O_{13}^2 =1$, 
such that 
\begin{align}
\left(\frac{\beta}{g_{L}}\right)^2
+\left(\frac{1}{g_X}\right)^2
+\left(\frac{1}{g_R}\right)^2=\frac{1}{g_Y^2}\ . 
\end{align}
Let us now consider an explicit representation for the orthogonal matrix $\mathcal{O}$ in terms 
of three angles $\omega$, $\phi$ and $\theta$, which are allowed to take values in the $[-\pi,\pi)$ interval. For convenience we choose
\begin{widetext}
\begin{align}
\mathcal{O} &=
\begin{pmatrix}
1 &0  &0  &0 \\
0 &\cosz &-\sinz &0 \\
0 &\sinz &\cosz &0 \\
0 &0 &0 &1
\end{pmatrix}
\begin{pmatrix}
1 &0  &0  &0 \\
0 &\cosy &0 &-\siny \\
0 &0 &1 &0 \\
0 &\siny &0 &\cosy
\end{pmatrix}
\begin{pmatrix}
1 &0  &0  &0 \\
0 &1 &0 &0 \\
0 &0 &\cosx &-\sinx \\
0 &0 &\sinx &\cosx
\end{pmatrix}\ .\notag 
\end{align}
\end{widetext}
From Eq.~\eqref{eq:angles1} we obtain 
\begin{align*}
\cos\phi \cos\omega= \frac{\beta g_Y}{g_{L}}\ ,
\hspace{0.3cm}
\cos\phi \sin\omega= \frac{ g_Y}{g_{X}}\ , 
\hspace{0.3cm}
\sin\phi = \frac{ g_Y}{g_{R}}\ , 
\end{align*}
From this relation we obtain
\begin{align*}
\cos\phi= \ha \frac{g_Y}{g_R}\ ,\hspace{0.3cm} 
\cos\omega=  \frac{\beta g_R}{\ha g_{L}}\ ,\hspace{0.3cm} 
\sin\omega=  \frac{ g_R}{\ha g_{X}}\ .
\end{align*}
 Here  
$\hat{\alpha}_{R} 
=g_R\sqrt{\frac{\beta^2}{g_{L}^2}+\frac{1}{g_X^2}}
=\sqrt{\frac{g_R^2}{g^2_L}\cot^2 \theta_W-1}$.
The typical left-right gauge coupling is $g_L=g_R=0.652$. 
By assuming 
\begin{align}
J_{Z'}=&\sum_f \bar{f}\left(\epsilon^{Z'}_L(f)P_L+\epsilon^{Z'}_R(f)P_R\right)f\ ,\notag\\
J_{8L}+J_{8R}=&\sum_f \bar{f}\left(\epsilon^{8}_{L}\ ,(f)P_L+\epsilon^{8}_R(f)P_R\right)f\ ,\notag\\
J_{3R}=&\sum_f \bar{f}\epsilon^{8}_R(f)P_Rf\ ,\notag\\
J_{X}=&\sum_f \bar{f}\left(X_{L}(f)P_L+X_R(f)P_R\right)f\ ,
\end{align}
where we define $\epsilon_{L,R}^a$ as the corresponding 
chiral charges associated with the gauge group generators $T_{L,R}^{a}=\frac{1}{2}\lambda^{a}$, with $\lambda^{a}$ 
the Gell-Mann matrices. 
Replacing  these expressions in Eq.~\eqref{Eq1b} and using  the elements of the rotation matrix $\mathcal{O}$  we get
the $Z'$ chiral charges: 
\begin{align}\label{eq:zprimecharges}
g_{Z^\prime}\epsilon^{Z^\prime}_{L,R}=&\ \ \ A_{L,R}\cosx+ B_{L,R}\sinx\ , \\
\gzpp       \epsilon^{\zpp}_{L,R}    =&-A_{L,R}\sinx+ B_{L,R}\cosx\ , 
 \end{align}
and $\theta$ is an angle  of the rotation matrix $\mathcal{O}$ which can take any value between 
$-\pi$ and $\pi$ and 
\begin{align}
 A_{L}=& 
  \hspace{0.3cm}
 \frac{g_R}{\ha}\left(
 \beta\dfrac{ g_X}{g_{L}}
 {X}_{L}
 -\hb  \epsilon^{8}_{L}\right) ,  \\
 A_{R}=& 
  \hspace{0.3cm}
 \frac{g_R}{\ha}\left(\beta\dfrac{ g_X}{g_{L}}
 {X}_{R}
 -\hb  \epsilon^{8}_{R} \right),  \\ 
B_{L}=&
    -\gp\dfrac{\beta \epsilon^{8}_{L}+X_L}{\ha}\ ,\\
  B_{R}=&
    \gp\left(-\dfrac{\beta \epsilon^{8}_{R}+ X_{R}}{\ha}+\ha \epsilon_{R}^{3}\right)\ .
\end{align}
In these expressions, to reproduce the SM charges it is necessary to take into account that $\beta (\epsilon^8_{L}+\epsilon^8_{R})+X=\frac{1}{2}(B-L)$, which is equivalent to 
$\beta \epsilon^8_{L,R}+X_{L,R}= \frac{1}{2}(B-L)$.
For exact left-right symmetry we have
$g_L=g_R$, replacing these identities in $A_{L,R}$ and $B_{L,R} $ we obtain the following. 
\begin{align}
 A_{L,R}=& 
  \hspace{0.3cm}
 \frac{g_L}{\ha}\left[
 \left(\dfrac{1}{z}+z\right)
 X_{L,R} 
 -\frac{z}{2}(B-L) \right] ,  \\
  B_{L}=&
    -\gp\dfrac{B-L}{2\ha}\ ,\\
  B_{R}=&
    \gp\left(-\dfrac{B-L}{2\ha}+\ha \epsilon_{R}^{3}\right)
    \ .
\end{align}
Here $z=\dfrac{g_{L}}{\beta g_X}
=\dfrac{1}{\beta}\sqrt{\cot^2\theta_W-\beta^2-1}=\dfrac{1}{\beta}\sqrt{\ha^2-\beta^2}$.
Under the previous assumptions with $\sin^2\theta_W= 0.23120$ 
(here, we use the value for the weak mixing angle in the $\overline{\text{MS}}$ scheme) we get $\ha=\sqrt{\cot^2 \theta_W-1}$. To avoid imaginary couplings, we require: $\beta<\ha\approx 1.525$, this condition rules out the notable case $\beta=\sqrt{3}$.
From these expressions, we obtain the chiral charges for the lepton and quark families.
These charges are reported in Appendix~\ref{sec:Azpcharges}.
\section{Low energy and collider constraints\label{sec:constraints}}

To establish lower bounds on the $Z'$ mass, we analyze results from searches for high-mass resonances decaying into dielectron and dimuon final states, focusing on the mass range 250\,GeV to 6\,TeV~\cite{ATLAS:2019erb}. These searches were performed using data collected by the ATLAS experiment during Run~2 of the Large Hadron Collider, corresponding to an integrated luminosity of 139\,fb$^{-1}$ at a center-of-mass energy of $\sqrt{s} = 13$\,TeV. The ATLAS Collaboration reported 95\% confidence level upper limits on the fiducial cross section times branching ratio, under various assumptions for the resonance width. For typical $E_6$-motivated couplings, these results translate into lower mass limits of approximately 4.5\,TeV.

In Figure~\ref{fig:colliders}, the lower limits on the $Z'$ mass are shown. 
To derive the constraints, we determine the value of the $Z'$ mass for which the QCD-predicted cross section matches the experimental upper limit on the differential cross section. 
To derive these limits, it is customary to assume that the masses of exotic fermions and right-handed neutrinos are effectively infinite. For a detailed description of the procedure and the corresponding theoretical expressions, we refer the reader to the references~\cite{Erler:2011ud,Salazar:2015gxa,Benavides:2018fzm}.
This approach applies for $Z'$ masses in the range 250~GeV to 6~TeV. Figure~\ref{fig:colliders} presents values of $M_{Z'}$ beyond 6~TeV. These should not be interpreted as lower bounds on the $Z'$ mass but rather as projections. The projections are obtained under the assumption that the upper limit on the cross section for $M_{Z'} > 6$~TeV remains below the current upper limit at 6~TeV. This assumption is well-motivated, as reduced background contributions are expected at higher masses, consistent with the typical behavior of exclusion limits in this regime.

\begin{widetext}
\begin{tabular}{cc}
 \includegraphics[scale=0.29]{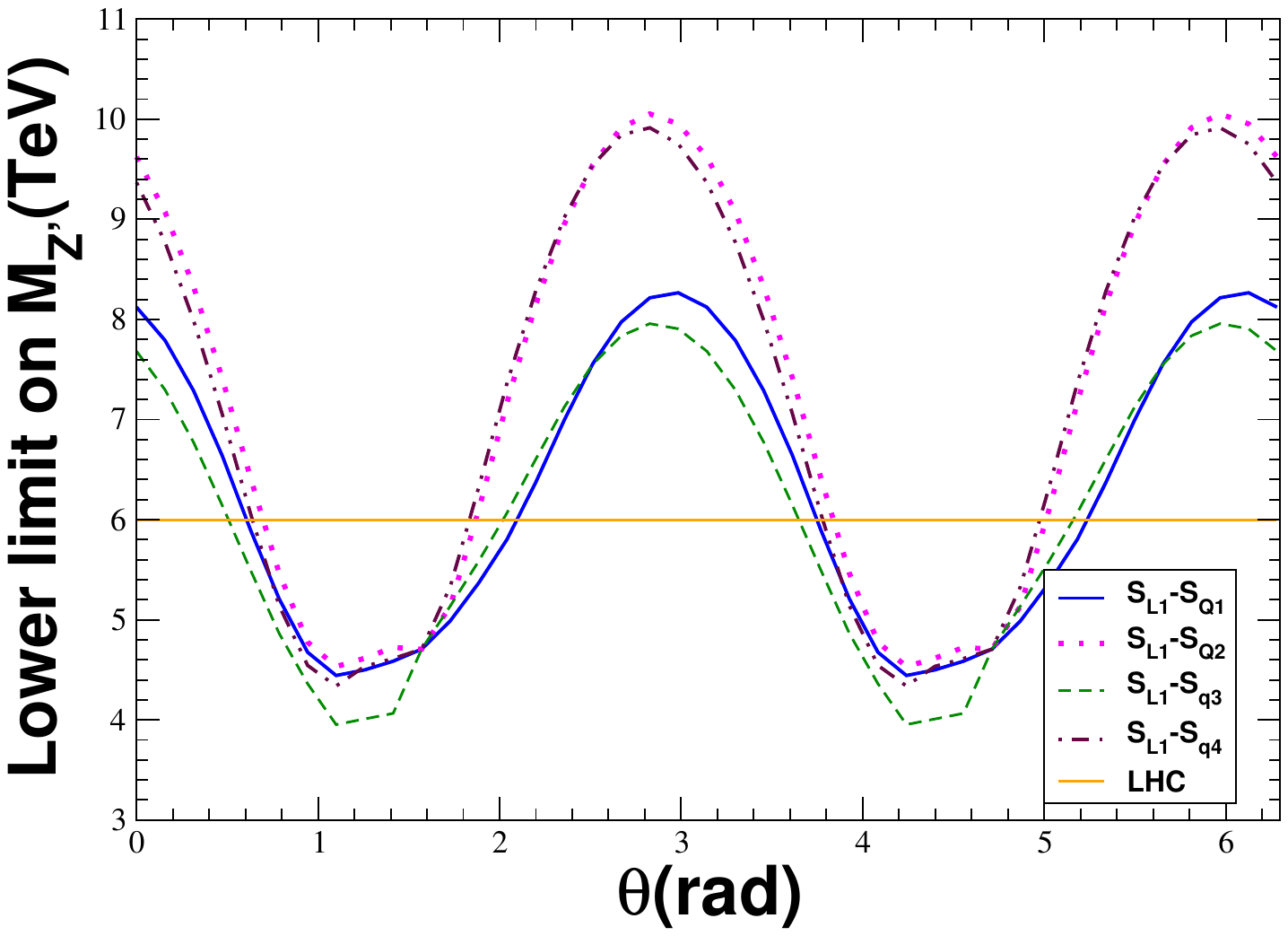}   &  \includegraphics[scale=0.29]{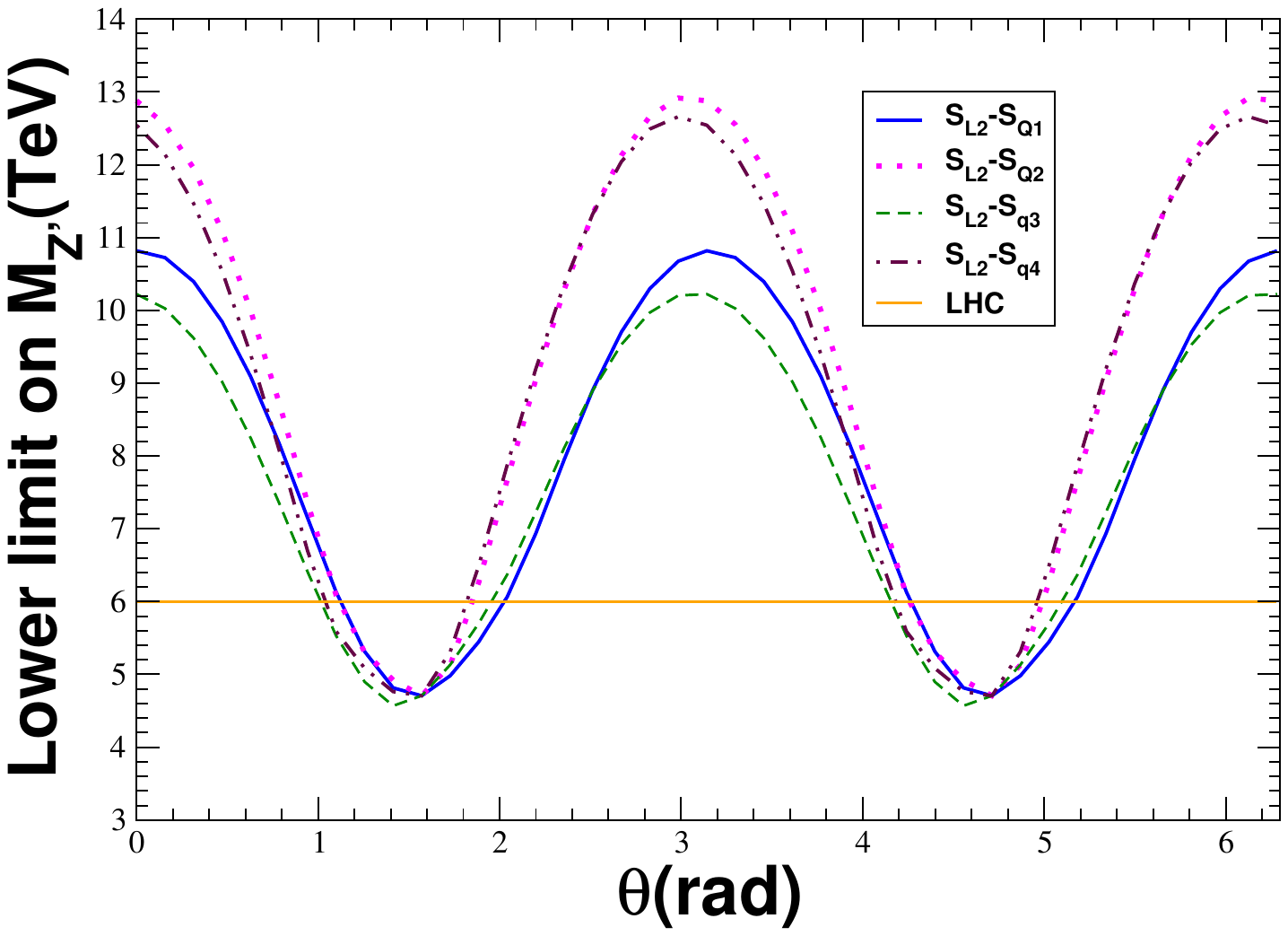}\\
\includegraphics[scale=0.29]{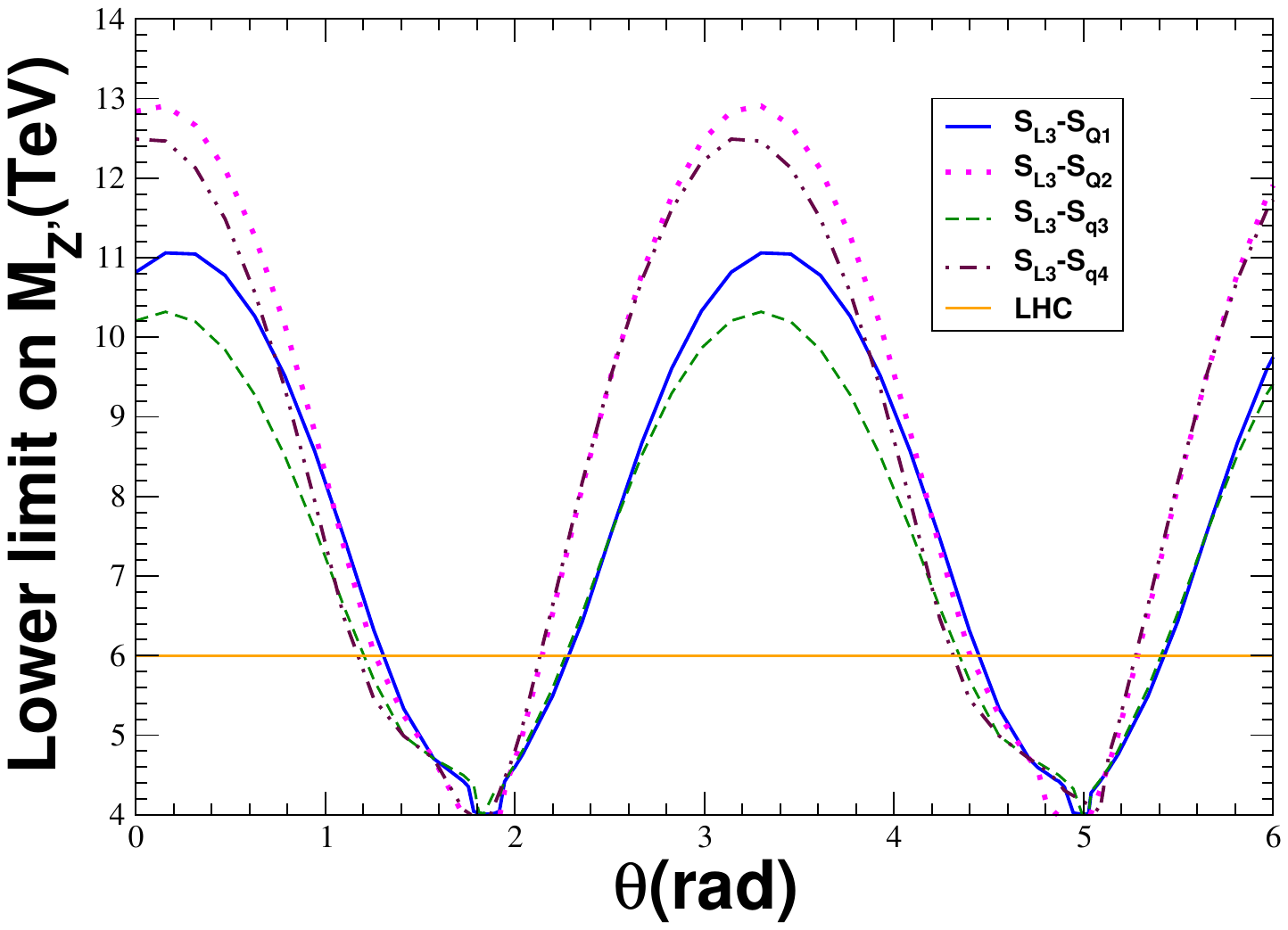}    &
\includegraphics[scale=0.29]{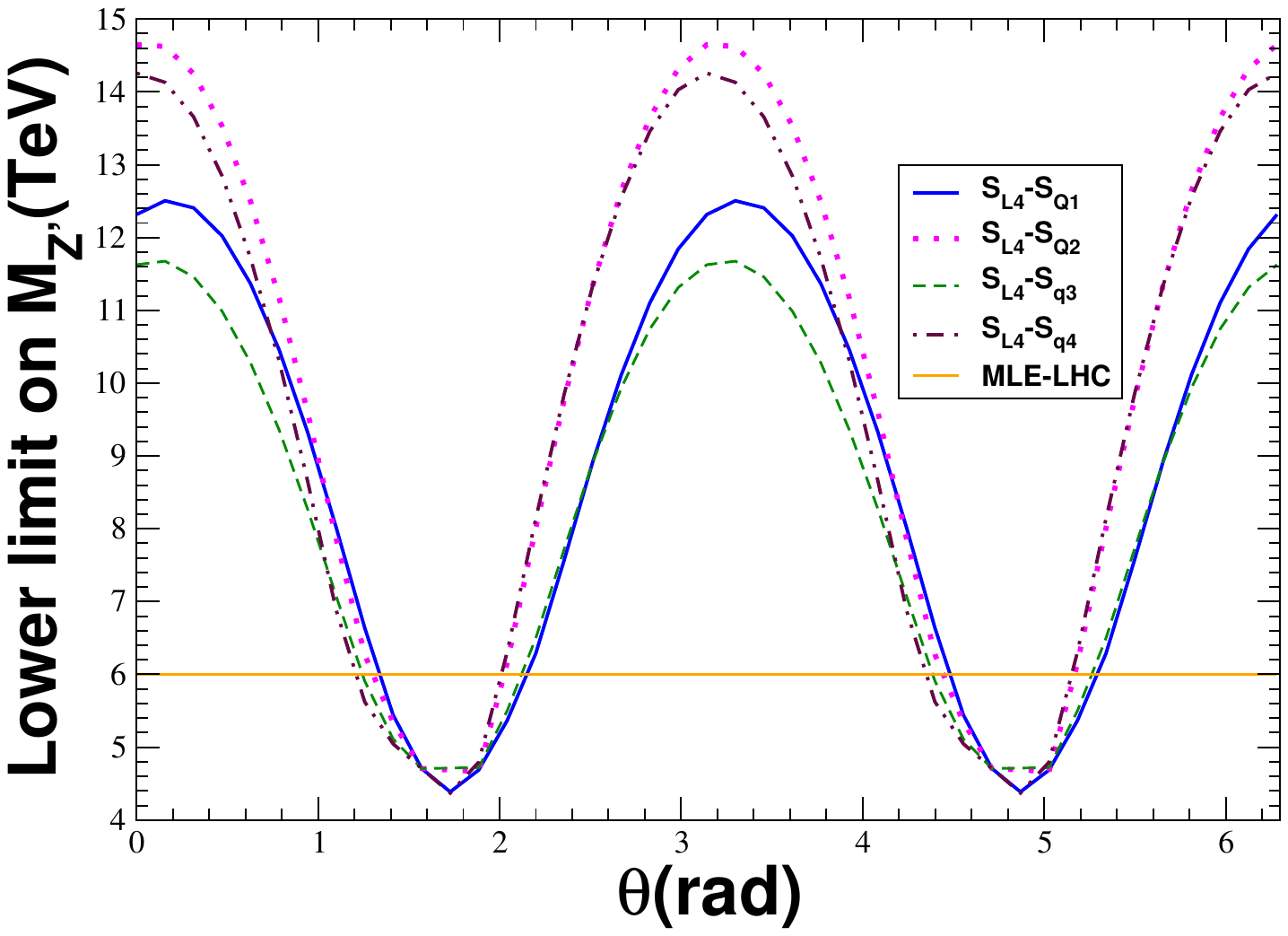} 
\end{tabular}

 \captionof{figure}{Left:  Lower limit on the $Z'$ mass.  We obtain these limits from the 95\%  CL upper limits on the fiducial $Z'$ production cross section times the $Z'  \rightarrow \ell^+\ell^-$ branching~\cite{ATLAS:2019erb}. The continuous yellow line labeled LHC refers to the highest $Z'$ mass for which the ATLAS Collaboration has reported upper limits on the production cross section.
 Restrictions for $Z'$ masses above 6 TeV are assumed to be projected limits. They are obtained assuming that the projected upper limit for the cross section in this region cannot exceed the ATLAS limit at 6 TeV~(The general trend of the upper limits on the cross section is to decrease with the mass of the $Z'$. Below this value, the ATLAS restrictions apply.
 $S_{Lj} - S_{Qi}$ indicates the chosen set of fermions for the $Z'$ charges of the first two families in the SM,  with $i, j = 1, 2, 3, 4$. The corresponding $Z'$ charges are shown in Appendix~\ref{sec:zpcharges}.
}
\label{fig:colliders}	
\end{widetext}

\section{Conclusions}

We have explored the extended gauge symmetry 
$SU(3)_C \otimes SU(3)_L \otimes SU(3)_R \otimes U(1)_X$,
which naturally embeds both the Left-Right and 3-3-1 models as subgroups. This embedding enables the construction of models that inherit the appealing features of these established frameworks.

Within this symmetry, we derived four lepton families and four quark families, and analyzed their contributions to gauge anomalies, as summarized in Table~\ref{tab:anomalias}. From these results, several anomaly-free sets of fermions were identified, as detailed in Section~\ref{sec:afs}. In total, we reported eight three-family and four two-family anomaly-free combinations for an arbitrary $\beta$.
The three-family anomaly-free ensembles are especially interesting, as they represent realistic extensions of the SM. These models are non-universal in the quark sector and universal in the lepton sector, inheriting many of the appealing features of the 3-3-1 models.
The two-family sets are useful for constructing models with an even number of families greater than or equal to four, offering flexibility in anomaly-free model building.

Finally, we derive constraints on the $Z'$ mass, accounting for the presence of a free parameter $\theta$ (a mixing angle) in the couplings to SM fermions. Our analysis reveals that collider bounds exhibit a strong dependence on the value of this parameter. These constraints were evaluated for all possible combinations of quark and lepton families, as illustrated in Figure~\ref{fig:colliders}.

Future work could investigate the high-energy unification of these models and their consequences for low-energy phenomenology. By introducing a suitable scalar potential, they may also serve as promising frameworks to study dark matter, neutrino properties, and CPT violation.

\begin{acknowledgments}
We thank W.~A.~Ponce for developing the formalism that underpins this work and for introducing us to this fascinating topic.
RB, ER, and YG acknowledge additional financial support from Minciencias CD82315 CT ICETEX 2021-1080. This research was partially supported by the \emph{Vicerrectoría de Investigaciones e Interacción Social}~(VIIS) de la
Universidad de Nariño, project numbers 2686, 2679, 3130, 3594 and 3595.
\end{acknowledgments}

\appendix

\section{$Z'$ charges \label{sec:Azpcharges}}
From Eq.~\eqref{eq:zprimecharges}, one can derive the $Z'$
 charges associated with each fermion family.
The $Z'$ charges for the Standard Model fields are presented in Tables~\ref{tab:sl1} through~\ref{tab:sq4}.
\begin{widetext}
\begin{center}
\begin{table}[h!]
\scalebox{0.8}{
\begin{tabular}{| c | c | c |}
\hline
\multicolumn{3}{|c|}{$S_{L1}$:  $\ell^T=(\nu_L,e_L)\subset \left(3,3,1,\frac{q-1}{3}\right)$, $(\nu_R,e_R)\subset \left(3,3,1,\frac{q-1}{3}\right)$  }\\
\hline
fields & $g_{Z'}\epsilon_{L}^{Z'}$ & $g_{Z'}\epsilon_{R}^{Z'}$ \\ \hline
$\nu_L$  
&$\frac{g_Y}{2}\left[\frac{1}{\ha}\right]s_{\theta}
+\frac{g_L}{\ha}\left[\frac{z}{2}+\frac{q-1}{3}(z+\frac{1}{z})\right]c_{\theta}$ & 
$\frac{g_Y}{2}\left[\frac{1}{\ha}+\ha\right]s_{\theta}
+\frac{g_L}{\ha}\left[\frac{z}{2}+\frac{q-1}{3}(z+\frac{1}{z})\right]c_{\theta}$
 \\ \hline
$e$ & $\frac{g_Y}{2}\left[\frac{1}{\ha}\right]s_{\theta}
+\frac{g_L}{\ha}\left[\frac{z}{2}+\frac{q-1}{3}(z+\frac{1}{z})\right]c_{\theta}$  
&
$\frac{g_Y}{2}\left[\frac{1}{\ha}-\ha\right]s_{\theta}
+\frac{g_L}{\ha}\left[\frac{z}{2}+\frac{q-1}{3}(z+\frac{1}{z})\right]c_{\theta}$ 
  \\ \hline
\hline
\end{tabular}
}\ 
\caption{$Z'$ chiral charges for the SM quarks when they are embedded in $S_{L1}$. 
Here $\ha=\sqrt{\cot^2 \theta_W-1}$  and $z=\dfrac{1}{\beta}\sqrt{\ha^2-\beta^2}$.
}
\label{tab:sl1}
\end{table}
\end{center}
\begin{center}
\begin{table}[h!]
\scalebox{0.8}{
\begin{tabular}{| c | c | c |}
\hline
\multicolumn{3}{|c|}{$S_{L2}$:  $\ell^T=(\nu_L,e_L)\subset \left(3,3,1,\frac{q-1}{3}\right)$, $(\nu_R,e_R)\subset \left(3,3^{*},1,-\frac{q+2}{3}\right)$  }\\
\hline
fields & $g_{Z'}\epsilon_{L}^{Z'}$ & $g_{Z'}\epsilon_{R}^{Z'}$ \\ \hline
$\nu_L$  
&$\frac{g_Y}{2}\left[\frac{1}{\ha}\right]s_{\theta}
+\frac{g_L}{\ha}\left[\frac{z}{2}+\frac{q-1}{3}(z+\frac{1}{z})\right]c_{\theta}$ & 
$\frac{g_Y}{2}\left[\frac{1}{\ha}+\ha\right]s_{\theta}
+\frac{g_L}{\ha}\left[\frac{z}{2}-\frac{2+q}{3}(z+\frac{1}{z})\right]c_{\theta}$
 \\ \hline
$e$ & $\frac{g_Y}{2}\left[\ha\right]s_{\theta}
+\frac{g_L}{\ha}\left[\frac{z}{2}+\frac{q-1}{3}(z+\frac{1}{z})\right]c_{\theta}$  
&
$\frac{g_Y}{2}\left[\frac{1}{\ha}-\ha\right]s_{\theta}
+\frac{g_L}{\ha}\left[\frac{z}{2}-\frac{2+q}{3}(z+\frac{1}{z})\right]c_{\theta}$ 
  \\ \hline
\hline
\end{tabular}
}\ 
\caption{$Z'$ chiral charges for the SM quarks when they are embedded in $S_{L2}$. Here $\ha=\sqrt{\cot^2 \theta_W-1}$  and $z=\dfrac{1}{\beta}\sqrt{\ha^2-\beta^2}$.}
\label{tab:sl2}
\end{table}
\end{center}
\begin{center}
\begin{table}[h!]
\scalebox{0.8}{
\begin{tabular}{| c | c | c |}
\hline
\multicolumn{3}{|c|}{$S_{L3}$:  $\ell^T=(\nu_L,e_L)\subset \left(3,3^{*},1,-\frac{q+2}{3}\right)$, $(\nu_R,e_R)\subset \left(3,3,1,\frac{q-1}{3}\right)$  }\\
\hline
fields & $g_{Z'}\epsilon_{L}^{Z'}$ & $g_{Z'}\epsilon_{R}^{Z'}$ \\ \hline
$\nu_L$  
&
$\frac{g_Y}{2}\left[\frac{1}{\ha}\right]s_{\theta}
+\frac{g_L}{\ha}\left[\frac{z}{2}-\frac{2+q}{3}(z+\frac{1}{z})\right]c_{\theta}$
& 
$\frac{g_Y}{2}\left[\frac{1}{\ha}+\ha\right]s_{\theta}
+\frac{g_L}{\ha}\left[\frac{z}{2}+\frac{q-1}{3}(z+\frac{1}{z})\right]c_{\theta}$ 
 \\ \hline
$e$ 
& 
$\frac{g_Y}{2}\left[\frac{1}{\ha}\right]s_{\theta}
+\frac{g_L}{\ha}\left[\frac{z}{2}-\frac{2+q}{3}(z+\frac{1}{z})\right]c_{\theta}$ 
&
$\frac{g_Y}{2}\left[\frac{1}{\ha}-\ha\right]s_{\theta}
+\frac{g_L}{\ha}\left[\frac{z}{2}+\frac{q-1}{3}(z+\frac{1}{z})\right]c_{\theta}$  
  \\ \hline
\hline
\end{tabular}
}\ 
\caption{$Z'$ chiral charges for the SM quarks when they are embedded in $S_{L3}$. Here $\ha=\sqrt{\cot^2 \theta_W-1}$  and $z=\dfrac{1}{\beta}\sqrt{\ha^2-\beta^2}$.}
\label{tab:sl3}
\end{table}
\end{center}    
\begin{center}
\begin{table}[h!]
\scalebox{0.8}{
\begin{tabular}{| c | c | c |}
\hline
\multicolumn{3}{|c|}{$S_{L4}$:  $\ell^T=(\nu_L,e_L)\subset \left(3,3^*,1,-\frac{q+2}{3}\right)$, $(\nu_R,e_R)\subset \left(3,3^*,1,-\frac{q+2}{3}\right)$  }\\
\hline
fields & $g_{Z'}\epsilon_{L}^{Z'}$ & $g_{Z'}\epsilon_{R}^{Z'}$ \\ \hline
$\nu_L$  
&
$\frac{g_Y}{2}\left[\frac{1}{\ha}\right]s_{\theta}
+\frac{g_L}{\ha}\left[\frac{z}{2}-\frac{2+q}{3}(z+\frac{1}{z})\right]c_{\theta}$
& 
$\frac{g_Y}{2}\left[\frac{1}{\ha}+\ha\right]s_{\theta}
+\frac{g_L}{\ha}\left[\frac{z}{2}-\frac{2+q}{3}(z+\frac{1}{z})\right]c_{\theta}$ 
 \\ \hline
$e$ 
& 
$\frac{g_Y}{2}\left[\frac{1}{\ha}\right]s_{\theta}
+\frac{g_L}{\ha}\left[\frac{z}{2}-\frac{2+q}{3}(z+\frac{1}{z})\right]c_{\theta}$ 
&
$\frac{g_Y}{2}\left[\ha-\frac{1}{\ha}\right]s_{\theta}
+\frac{g_L}{\ha}\left[\frac{z}{2}-\frac{2+q}{3}(z+\frac{1}{z})\right]c_{\theta}$  
  \\ \hline
\hline
\end{tabular}
}\ 
\caption{$Z'$ chiral charges for the SM quarks when they are embedded in $S_{L4}$. Here $\ha=\sqrt{\cot^2 \theta_W-1}$  and $z=\dfrac{1}{\beta}\sqrt{\ha^2-\beta^2}$.}
\label{tab:sl4}
\end{table}
\end{center}    
\begin{center}
\begin{table}[h!]
\scalebox{0.8}{
\begin{tabular}{| c | c | c |}
\hline
\multicolumn{3}{|c|}{$S_{Q1}$:  $q^T=(u_L,d_L)\subset \left(3,3,1,\frac{q+1}{3}\right)$, $(u_R,d_R)\subset \left(3,3,1,\frac{q+1}{3}\right)$  }\\
\hline
fields & $g_{Z'}\epsilon_{L}^{Z'}$ & $g_{Z'}\epsilon_{R}^{Z'}$ \\ \hline
$u$  & $-g_Y\left[\frac{1}{6\ha}\right]s_{\theta}
+\frac{g_L}{3\ha}\left[-\frac{z}{2}+(1+q)(z+\frac{1}{z})\right]c_{\theta}
$ & $-g_Y\left[\frac{1}{6\ha}-\frac{\ha}{2}\right]s_{\theta}
+\frac{g_L}{3\ha}\left[-\frac{z}{2}+(1+q)(z+\frac{1}{z})\right]c_{\theta}
$ \\ \hline
$d$ & $-g_Y\left[\frac{1}{6\ha}\right]s_{\theta}
+\frac{g_L}{3\ha}\left[-\frac{z}{2}+(1+q)(z+\frac{1}{z})\right]c_{\theta}
$ & $-g_Y\left[\frac{1}{6\ha}+\frac{\ha}{2}\right]s_{\theta}
+\frac{g_L}{3\ha}\left[-\frac{z}{2}+(1+q)(z+\frac{1}{z})\right]c_{\theta}
$  \\ \hline
\hline
\end{tabular}
}\ 
\caption{$Z'$ chiral charges for the SM quarks when they are embedded in $S_{Q1}$. Here $\ha=\sqrt{\cot^2 \theta_W-1}$  and $z=\dfrac{1}{\beta}\sqrt{\ha^2-\beta^2}$.}
\label{tab:sq1}
\end{table}
\end{center}
\begin{center}
\begin{table}[h!]
\scalebox{0.8}{
\begin{tabular}{| c | c | c |}
\hline
\multicolumn{3}{|c|}{$S_{Q2}$:  $q^T=(u_L,d_L)\subset \left(3,3,1,\frac{q+1}{3}\right)$, $(u_R,d_R)\subset \left(3,3^*,1,-\frac{q}{3}\right)$  }\\
\hline
fields & $g_{Z'}\epsilon_{L}^{Z'}$ & $g_{Z'}\epsilon_{R}^{Z'}$ \\ \hline
$u$  & $-\frac{g_Y}{6\ha}s_{\theta}
+\frac{g_L}{3\ha}\left[-\frac{z}{2}+(1+q)(z+\frac{1}{z})\right]c_{\theta}$
& $-g_Y\left[\frac{1}{6\ha}-\frac{\ha}{2}\right]s_{\theta}
-\frac{g_L}{3\ha}\left[\frac{z}{2}+q(z+\frac{1}{z})\right]c_{\theta}
$ \\ \hline
$d$ & $-\frac{g_Y}{6\ha}s_{\theta}
+\frac{g_L}{3\ha}\left[-\frac{z}{2}+(1+q)(z+\frac{1}{z})\right]c_{\theta}
$ & $-g_Y\left[\frac{\ha}{2}+\frac{1}{6\ha}\right]s_{\theta}
-\frac{g_L}{3\ha}\left[\frac{z}{2}+q(z+\frac{1}{z})\right]c_{\theta}
$  \\ \hline
\hline
\end{tabular}
}\ 
\caption{$Z'$ chiral charges for the SM quarks when they are embedded in $S_{Q2}$. Here $\ha=\sqrt{\cot^2 \theta_W-1}$  and $z=\dfrac{1}{\beta}\sqrt{\ha^2-\beta^2}$.}
\label{tab:sq2}
\end{table}
\end{center}
\begin{center}
\begin{table}[h!]
\scalebox{0.8}{
\begin{tabular}{| c | c | c |}
\hline
\multicolumn{3}{|c|}{$S_{Q3}$:  $q^T=(u_L,d_L)\subset \left(3,3^*,1,-\frac{q}{3}\right)$, $(u_R,d_R)\subset \left(3,3,1,\frac{1+q}{3}\right)$  }\\
\hline
fields & $g_{Z'}\epsilon_{L}^{Z'}$ & $g_{Z'}\epsilon_{R}^{Z'}$ \\ \hline
$u$ 
& $-\frac{g_Y}{6\ha}s_{\theta}
-\frac{g_L}{3\ha}\left[\frac{z}{2}+q(z+\frac{1}{z})\right]c_{\theta}
$ 
& $g_Y\left[\frac{\ha}{2}-\frac{1}{6\ha}\right]s_{\theta}
+\frac{g_L}{3\ha}\left[-\frac{z}{2}+(1+q)(z+\frac{1}{z})\right]c_{\theta}$\\
 \hline
$d$ 
& $-\frac{g_Y}{6\ha}s_{\theta}
-\frac{g_L}{3\ha}\left[\frac{z}{2}+q(z+\frac{1}{z})\right]c_{\theta}$
& $-g_Y\left[\frac{\ha}{2}+\frac{1}{6\ha}\right]s_{\theta}
+\frac{g_L}{3\ha}\left[-\frac{z}{2}+(1+q)(z+\frac{1}{z})\right]c_{\theta}$ 
  \\ \hline
\hline
\end{tabular}
}\ 
\caption{$Z'$ chiral charges for the SM quarks when they are embedded in $S_{Q3}$. Here $\ha=\sqrt{\cot^2 \theta_W-1}$  and $z=\dfrac{1}{\beta}\sqrt{\ha^2-\beta^2}$.}
\label{tab:sq3}
\end{table}
\end{center}
\begin{center}
\begin{table}[h!]
\scalebox{0.8}{
\begin{tabular}{| c | c | c |}
\hline
\multicolumn{3}{|c|}{$S_{Q4}$:  $q^T=(u_L,d_L)\subset \left(3,3^*,1,-\frac{q}{3}\right)$, $(u_R,d_R)\subset \left(3,3^*,1,-\frac{q}{3}\right)$  }\\
\hline
fields & $g_{Z'}\epsilon_{L}^{Z'}$ & $g_{Z'}\epsilon_{R}^{Z'}$ \\ \hline
$u$ 
& $-\frac{g_Y}{6\ha}s_{\theta}
-\frac{g_L}{3\ha}\left[\frac{z}{2}+q(z+\frac{1}{z})\right]c_{\theta}
$ 
& $g_Y\left[\frac{\ha}{2}-\frac{1}{6\ha}\right]s_{\theta}
-\frac{g_L}{3\ha}\left[\frac{z}{2}+q(z+\frac{1}{z})\right]c_{\theta}$\\
 \hline
$d$ 
& $-\frac{1}{6\ha}s_{\theta}
-\frac{g_L}{3\ha}\left[\frac{z}{2}+q(z+\frac{1}{z})\right]c_{\theta}$
& $-g_Y\left[\frac{\ha}{2}+\frac{1}{6\ha}\right]s_{\theta}
-\frac{g_L}{3\ha}\left[\frac{z}{2}+q(z+\frac{1}{z})\right]c_{\theta}$ 
  \\ \hline
\hline
\end{tabular}
}\ 
\caption{$Z'$ chiral charges for the SM quarks when they are embedded in $S_{Q4}$.Here $\ha=\sqrt{\cot^2 \theta_W-1}$  and $z=\dfrac{1}{\beta}\sqrt{\ha^2-\beta^2}$.}
\label{tab:sq4}
\end{table}
\end{center}

\end{widetext}



\nocite{*}

\bibliography{aipsamp}

\providecommand{\noopsort}[1]{}\providecommand{\singleletter}[1]{#1}%
\begin{thebibliography}{20}%
\makeatletter
\providecommand \@ifxundefined [1]{%
 \@ifx{#1\undefined}
}%
\providecommand \@ifnum [1]{%
 \ifnum #1\expandafter \@firstoftwo
 \else \expandafter \@secondoftwo
 \fi
}%
\providecommand \@ifx [1]{%
 \ifx #1\expandafter \@firstoftwo
 \else \expandafter \@secondoftwo
 \fi
}%
\providecommand \natexlab [1]{#1}%
\providecommand \enquote  [1]{``#1''}%
\providecommand \bibnamefont  [1]{#1}%
\providecommand \bibfnamefont [1]{#1}%
\providecommand \citenamefont [1]{#1}%
\providecommand \href@noop [0]{\@secondoftwo}%
\providecommand \href [0]{\begingroup \@sanitize@url \@href}%
\providecommand \@href[1]{\@@startlink{#1}\@@href}%
\providecommand \@@href[1]{\endgroup#1\@@endlink}%
\providecommand \@sanitize@url [0]{\catcode `\\12\catcode `\$12\catcode
  `\&12\catcode `\#12\catcode `\^12\catcode `\_12\catcode `\%12\relax}%
\providecommand \@@startlink[1]{}%
\providecommand \@@endlink[0]{}%
\providecommand \url  [0]{\begingroup\@sanitize@url \@url }%
\providecommand \@url [1]{\endgroup\@href {#1}{\urlprefix }}%
\providecommand \urlprefix  [0]{URL }%
\providecommand \Eprint [0]{\href }%
\providecommand \doibase [0]{https://doi.org/}%
\providecommand \selectlanguage [0]{\@gobble}%
\providecommand \bibinfo  [0]{\@secondoftwo}%
\providecommand \bibfield  [0]{\@secondoftwo}%
\providecommand \translation [1]{[#1]}%
\providecommand \BibitemOpen [0]{}%
\providecommand \bibitemStop [0]{}%
\providecommand \bibitemNoStop [0]{.\EOS\space}%
\providecommand \EOS [0]{\spacefactor3000\relax}%
\providecommand \BibitemShut  [1]{\csname bibitem#1\endcsname}%
\let\auto@bib@innerbib\@empty
\bibitem [{\citenamefont {Foot}\ \emph {et~al.}(1993)\citenamefont {Foot},
  \citenamefont {Lew},\ and\ \citenamefont {Volkas}}]{Foot:1992ui}%
  \BibitemOpen
  \bibfield  {author} {\bibinfo {author} {\bibfnamefont {R.}~\bibnamefont
  {Foot}}, \bibinfo {author} {\bibfnamefont {H.}~\bibnamefont {Lew}},\ and\
  \bibinfo {author} {\bibfnamefont {R.~R.}\ \bibnamefont {Volkas}},\ }\bibfield
   {title} {\bibinfo {title} {{Electric charge quantization}},\ }\href
  {https://doi.org/10.1088/0954-3899/19/3/005} {\bibfield  {journal} {\bibinfo
  {journal} {J. Phys. G}\ }\textbf {\bibinfo {volume} {19}},\ \bibinfo {pages}
  {361} (\bibinfo {year} {1993})},\ \bibinfo {note} {[Erratum: J.Phys.G 19,
  1067 (1993)]},\ \Eprint {https://arxiv.org/abs/hep-ph/9209259}
  {arXiv:hep-ph/9209259} \BibitemShut {NoStop}%
\bibitem [{\citenamefont {Nowakowski}\ and\ \citenamefont
  {Pilaftsis}(1993)}]{Nowakowski:1992ff}%
  \BibitemOpen
  \bibfield  {author} {\bibinfo {author} {\bibfnamefont {M.}~\bibnamefont
  {Nowakowski}}\ and\ \bibinfo {author} {\bibfnamefont {A.}~\bibnamefont
  {Pilaftsis}},\ }\bibfield  {title} {\bibinfo {title} {{A Note on charge
  quantization through anomaly cancellation}},\ }\href
  {https://doi.org/10.1103/PhysRevD.48.259} {\bibfield  {journal} {\bibinfo
  {journal} {Phys. Rev. D}\ }\textbf {\bibinfo {volume} {48}},\ \bibinfo
  {pages} {259} (\bibinfo {year} {1993})},\ \Eprint
  {https://arxiv.org/abs/hep-ph/9304312} {arXiv:hep-ph/9304312} \BibitemShut
  {NoStop}%
\bibitem [{\citenamefont {Pati}\ and\ \citenamefont
  {Salam}(1974)}]{Pati:1974yy}%
  \BibitemOpen
  \bibfield  {author} {\bibinfo {author} {\bibfnamefont {J.~C.}\ \bibnamefont
  {Pati}}\ and\ \bibinfo {author} {\bibfnamefont {A.}~\bibnamefont {Salam}},\
  }\bibfield  {title} {\bibinfo {title} {{Lepton Number as the Fourth Color}},\
  }\href {https://doi.org/10.1103/PhysRevD.10.275} {\bibfield  {journal}
  {\bibinfo  {journal} {Phys. Rev. D}\ }\textbf {\bibinfo {volume} {10}},\
  \bibinfo {pages} {275} (\bibinfo {year} {1974})},\ \bibinfo {note} {[Erratum:
  Phys.Rev.D 11, 703--703 (1975)]}\BibitemShut {NoStop}%
\bibitem [{\citenamefont {Georgi}\ and\ \citenamefont
  {Glashow}(1974)}]{Georgi:1974sy}%
  \BibitemOpen
  \bibfield  {author} {\bibinfo {author} {\bibfnamefont {H.}~\bibnamefont
  {Georgi}}\ and\ \bibinfo {author} {\bibfnamefont {S.~L.}\ \bibnamefont
  {Glashow}},\ }\bibfield  {title} {\bibinfo {title} {{Unity of All Elementary
  Particle Forces}},\ }\href {https://doi.org/10.1103/PhysRevLett.32.438}
  {\bibfield  {journal} {\bibinfo  {journal} {Phys. Rev. Lett.}\ }\textbf
  {\bibinfo {volume} {32}},\ \bibinfo {pages} {438} (\bibinfo {year}
  {1974})}\BibitemShut {NoStop}%
\bibitem [{\citenamefont {Reig}\ \emph {et~al.}(2017)\citenamefont {Reig},
  \citenamefont {Valle},\ and\ \citenamefont {Vaquera-Araujo}}]{Reig:2016tuk}%
  \BibitemOpen
  \bibfield  {author} {\bibinfo {author} {\bibfnamefont {M.}~\bibnamefont
  {Reig}}, \bibinfo {author} {\bibfnamefont {J.~W.~F.}\ \bibnamefont {Valle}},\
  and\ \bibinfo {author} {\bibfnamefont {C.~A.}\ \bibnamefont
  {Vaquera-Araujo}},\ }\bibfield  {title} {\bibinfo {title} {{Unifying
  left\textendash{}right symmetry and 331 electroweak theories}},\ }\href
  {https://doi.org/10.1016/j.physletb.2016.12.049} {\bibfield  {journal}
  {\bibinfo  {journal} {Phys. Lett. B}\ }\textbf {\bibinfo {volume} {766}},\
  \bibinfo {pages} {35} (\bibinfo {year} {2017})},\ \Eprint
  {https://arxiv.org/abs/1611.02066} {arXiv:1611.02066 [hep-ph]} \BibitemShut
  {NoStop}%
\bibitem [{\citenamefont {Hati}\ \emph {et~al.}(2017)\citenamefont {Hati},
  \citenamefont {Patra}, \citenamefont {Reig}, \citenamefont {Valle},\ and\
  \citenamefont {Vaquera-Araujo}}]{Hati:2017aez}%
  \BibitemOpen
  \bibfield  {author} {\bibinfo {author} {\bibfnamefont {C.}~\bibnamefont
  {Hati}}, \bibinfo {author} {\bibfnamefont {S.}~\bibnamefont {Patra}},
  \bibinfo {author} {\bibfnamefont {M.}~\bibnamefont {Reig}}, \bibinfo {author}
  {\bibfnamefont {J.~W.~F.}\ \bibnamefont {Valle}},\ and\ \bibinfo {author}
  {\bibfnamefont {C.~A.}\ \bibnamefont {Vaquera-Araujo}},\ }\bibfield  {title}
  {\bibinfo {title} {{Towards gauge coupling unification in left-right
  symmetric $\mathrm{SU(3)_c \times SU(3)_L \times SU(3)_R \times U(1)_{X}}$
  theories}},\ }\href {https://doi.org/10.1103/PhysRevD.96.015004} {\bibfield
  {journal} {\bibinfo  {journal} {Phys. Rev. D}\ }\textbf {\bibinfo {volume}
  {96}},\ \bibinfo {pages} {015004} (\bibinfo {year} {2017})},\ \Eprint
  {https://arxiv.org/abs/1703.09647} {arXiv:1703.09647 [hep-ph]} \BibitemShut
  {NoStop}%
\bibitem [{\citenamefont {Dong}\ \emph {et~al.}(2018)\citenamefont {Dong},
  \citenamefont {Huong}, \citenamefont {Queiroz}, \citenamefont {Valle},\ and\
  \citenamefont {Vaquera-Araujo}}]{Dong:2017zxo}%
  \BibitemOpen
  \bibfield  {author} {\bibinfo {author} {\bibfnamefont {P.~V.}\ \bibnamefont
  {Dong}}, \bibinfo {author} {\bibfnamefont {D.~T.}\ \bibnamefont {Huong}},
  \bibinfo {author} {\bibfnamefont {F.~S.}\ \bibnamefont {Queiroz}}, \bibinfo
  {author} {\bibfnamefont {J.~W.~F.}\ \bibnamefont {Valle}},\ and\ \bibinfo
  {author} {\bibfnamefont {C.~A.}\ \bibnamefont {Vaquera-Araujo}},\ }\bibfield
  {title} {\bibinfo {title} {{The Dark Side of Flipped Trinification}},\ }\href
  {https://doi.org/10.1007/JHEP04(2018)143} {\bibfield  {journal} {\bibinfo
  {journal} {JHEP}\ }\textbf {\bibinfo {volume} {04}},\ \bibinfo {pages}
  {143}},\ \Eprint {https://arxiv.org/abs/1710.06951} {arXiv:1710.06951
  [hep-ph]} \BibitemShut {NoStop}%
\bibitem [{\citenamefont {Huong}\ \emph {et~al.}(2018)\citenamefont {Huong},
  \citenamefont {Dong}, \citenamefont {Duy}, \citenamefont {Nhuan},\ and\
  \citenamefont {Thien}}]{Huong:2018ytz}%
  \BibitemOpen
  \bibfield  {author} {\bibinfo {author} {\bibfnamefont {D.~T.}\ \bibnamefont
  {Huong}}, \bibinfo {author} {\bibfnamefont {P.~V.}\ \bibnamefont {Dong}},
  \bibinfo {author} {\bibfnamefont {N.~T.}\ \bibnamefont {Duy}}, \bibinfo
  {author} {\bibfnamefont {N.~T.}\ \bibnamefont {Nhuan}},\ and\ \bibinfo
  {author} {\bibfnamefont {L.~D.}\ \bibnamefont {Thien}},\ }\bibfield  {title}
  {\bibinfo {title} {{Investigation of Dark Matter in the 3-2-3-1 Model}},\
  }\href {https://doi.org/10.1103/PhysRevD.98.055033} {\bibfield  {journal}
  {\bibinfo  {journal} {Phys. Rev. D}\ }\textbf {\bibinfo {volume} {98}},\
  \bibinfo {pages} {055033} (\bibinfo {year} {2018})},\ \Eprint
  {https://arxiv.org/abs/1802.10402} {arXiv:1802.10402 [hep-ph]} \BibitemShut
  {NoStop}%
\bibitem [{\citenamefont {Dinh}\ \emph {et~al.}(2019)\citenamefont {Dinh},
  \citenamefont {Huong}, \citenamefont {Duy}, \citenamefont {Nhuan},
  \citenamefont {Thien},\ and\ \citenamefont {Van~Dong}}]{Dinh:2019jdg}%
  \BibitemOpen
  \bibfield  {author} {\bibinfo {author} {\bibfnamefont {D.~N.}\ \bibnamefont
  {Dinh}}, \bibinfo {author} {\bibfnamefont {D.~T.}\ \bibnamefont {Huong}},
  \bibinfo {author} {\bibfnamefont {N.~T.}\ \bibnamefont {Duy}}, \bibinfo
  {author} {\bibfnamefont {N.~T.}\ \bibnamefont {Nhuan}}, \bibinfo {author}
  {\bibfnamefont {L.~D.}\ \bibnamefont {Thien}},\ and\ \bibinfo {author}
  {\bibfnamefont {P.}~\bibnamefont {Van~Dong}},\ }\bibfield  {title} {\bibinfo
  {title} {{Flavor changing in the flipped trinification}},\ }\href
  {https://doi.org/10.1103/PhysRevD.99.055005} {\bibfield  {journal} {\bibinfo
  {journal} {Phys. Rev. D}\ }\textbf {\bibinfo {volume} {99}},\ \bibinfo
  {pages} {055005} (\bibinfo {year} {2019})},\ \Eprint
  {https://arxiv.org/abs/1901.07969} {arXiv:1901.07969 [hep-ph]} \BibitemShut
  {NoStop}%
\bibitem [{\citenamefont {Singer}\ \emph {et~al.}(1980)\citenamefont {Singer},
  \citenamefont {Valle},\ and\ \citenamefont {Schechter}}]{Singer:1980sw}%
  \BibitemOpen
  \bibfield  {author} {\bibinfo {author} {\bibfnamefont {M.}~\bibnamefont
  {Singer}}, \bibinfo {author} {\bibfnamefont {J.~W.~F.}\ \bibnamefont
  {Valle}},\ and\ \bibinfo {author} {\bibfnamefont {J.}~\bibnamefont
  {Schechter}},\ }\bibfield  {title} {\bibinfo {title} {{Canonical Neutral
  Current Predictions From the Weak Electromagnetic Gauge Group SU(3) X
  $u$(1)}},\ }\href {https://doi.org/10.1103/PhysRevD.22.738} {\bibfield
  {journal} {\bibinfo  {journal} {Phys. Rev. D}\ }\textbf {\bibinfo {volume}
  {22}},\ \bibinfo {pages} {738} (\bibinfo {year} {1980})}\BibitemShut
  {NoStop}%
\bibitem [{\citenamefont {Pisano}\ and\ \citenamefont
  {Pleitez}(1992)}]{Pisano:1992bxx}%
  \BibitemOpen
  \bibfield  {author} {\bibinfo {author} {\bibfnamefont {F.}~\bibnamefont
  {Pisano}}\ and\ \bibinfo {author} {\bibfnamefont {V.}~\bibnamefont
  {Pleitez}},\ }\bibfield  {title} {\bibinfo {title} {{An SU(3) x U(1) model
  for electroweak interactions}},\ }\href
  {https://doi.org/10.1103/PhysRevD.46.410} {\bibfield  {journal} {\bibinfo
  {journal} {Phys. Rev. D}\ }\textbf {\bibinfo {volume} {46}},\ \bibinfo
  {pages} {410} (\bibinfo {year} {1992})},\ \Eprint
  {https://arxiv.org/abs/hep-ph/9206242} {arXiv:hep-ph/9206242} \BibitemShut
  {NoStop}%
\bibitem [{\citenamefont {Frampton}(1992)}]{Frampton:1992wt}%
  \BibitemOpen
  \bibfield  {author} {\bibinfo {author} {\bibfnamefont {P.~H.}\ \bibnamefont
  {Frampton}},\ }\bibfield  {title} {\bibinfo {title} {{Chiral dilepton model
  and the flavor question}},\ }\href
  {https://doi.org/10.1103/PhysRevLett.69.2889} {\bibfield  {journal} {\bibinfo
   {journal} {Phys. Rev. Lett.}\ }\textbf {\bibinfo {volume} {69}},\ \bibinfo
  {pages} {2889} (\bibinfo {year} {1992})}\BibitemShut {NoStop}%
\bibitem [{\citenamefont {Benavides}\ \emph {et~al.}(2022)\citenamefont
  {Benavides}, \citenamefont {Giraldo}, \citenamefont {Mu\~noz}, \citenamefont
  {Ponce},\ and\ \citenamefont {Rojas}}]{Benavides:2021pqx}%
  \BibitemOpen
  \bibfield  {author} {\bibinfo {author} {\bibfnamefont {R.~H.}\ \bibnamefont
  {Benavides}}, \bibinfo {author} {\bibfnamefont {Y.}~\bibnamefont {Giraldo}},
  \bibinfo {author} {\bibfnamefont {L.}~\bibnamefont {Mu\~noz}}, \bibinfo
  {author} {\bibfnamefont {W.~A.}\ \bibnamefont {Ponce}},\ and\ \bibinfo
  {author} {\bibfnamefont {E.}~\bibnamefont {Rojas}},\ }\bibfield  {title}
  {\bibinfo {title} {{Systematic study of the SU(3)$_{c}$
  \ensuremath{\otimes}SU(3)$_{L}$ \ensuremath{\otimes} U(1)$_{X}$ local gauge
  symmetry}},\ }\href {https://doi.org/10.1088/1361-6471/ac894a} {\bibfield
  {journal} {\bibinfo  {journal} {J. Phys. G}\ }\textbf {\bibinfo {volume}
  {49}},\ \bibinfo {pages} {105007} (\bibinfo {year} {2022})},\ \Eprint
  {https://arxiv.org/abs/2111.02563} {arXiv:2111.02563 [hep-ph]} \BibitemShut
  {NoStop}%
\bibitem [{\citenamefont {Suarez}\ \emph {et~al.}(2024)\citenamefont {Suarez},
  \citenamefont {Benavides}, \citenamefont {Giraldo}, \citenamefont {Ponce},\
  and\ \citenamefont {Rojas}}]{Suarez:2023ozu}%
  \BibitemOpen
  \bibfield  {author} {\bibinfo {author} {\bibfnamefont {E.}~\bibnamefont
  {Suarez}}, \bibinfo {author} {\bibfnamefont {R.~H.}\ \bibnamefont
  {Benavides}}, \bibinfo {author} {\bibfnamefont {Y.}~\bibnamefont {Giraldo}},
  \bibinfo {author} {\bibfnamefont {W.~A.}\ \bibnamefont {Ponce}},\ and\
  \bibinfo {author} {\bibfnamefont {E.}~\bibnamefont {Rojas}},\ }\bibfield
  {title} {\bibinfo {title} {{Alternative 3-3-1 models with exotic electric
  charges}},\ }\href {https://doi.org/10.1088/1361-6471/ad1e21} {\bibfield
  {journal} {\bibinfo  {journal} {J. Phys. G}\ }\textbf {\bibinfo {volume}
  {51}},\ \bibinfo {pages} {035004} (\bibinfo {year} {2024})},\ \Eprint
  {https://arxiv.org/abs/2307.15826} {arXiv:2307.15826 [hep-ph]} \BibitemShut
  {NoStop}%
\bibitem [{\citenamefont {Mohapatra}\ and\ \citenamefont
  {Pati}(1975)}]{Mohapatra:1974hk}%
  \BibitemOpen
  \bibfield  {author} {\bibinfo {author} {\bibfnamefont {R.~N.}\ \bibnamefont
  {Mohapatra}}\ and\ \bibinfo {author} {\bibfnamefont {J.~C.}\ \bibnamefont
  {Pati}},\ }\bibfield  {title} {\bibinfo {title} {{Left-Right Gauge Symmetry
  and an Isoconjugate Model of CP Violation}},\ }\href
  {https://doi.org/10.1103/PhysRevD.11.566} {\bibfield  {journal} {\bibinfo
  {journal} {Phys. Rev. D}\ }\textbf {\bibinfo {volume} {11}},\ \bibinfo
  {pages} {566} (\bibinfo {year} {1975})}\BibitemShut {NoStop}%
\bibitem [{\citenamefont {Franceschini}(2023)}]{Franceschini:2023nlp}%
  \BibitemOpen
  \bibfield  {author} {\bibinfo {author} {\bibfnamefont {R.}~\bibnamefont
  {Franceschini}},\ }\bibfield  {title} {\bibinfo {title} {{Physics Beyond the
  Standard Model Associated with the Top Quark}},\ }\href
  {https://doi.org/10.1146/annurev-nucl-102020-011427} {\bibfield  {journal}
  {\bibinfo  {journal} {Ann. Rev. Nucl. Part. Sci.}\ }\textbf {\bibinfo
  {volume} {73}},\ \bibinfo {pages} {397} (\bibinfo {year} {2023})},\ \Eprint
  {https://arxiv.org/abs/2301.04407} {arXiv:2301.04407 [hep-ph]} \BibitemShut
  {NoStop}%
\bibitem [{\citenamefont {Aad}\ \emph {et~al.}(2019)\citenamefont {Aad} \emph
  {et~al.}}]{ATLAS:2019erb}%
  \BibitemOpen
  \bibfield  {author} {\bibinfo {author} {\bibfnamefont {G.}~\bibnamefont
  {Aad}} \emph {et~al.} (\bibinfo {collaboration} {ATLAS}),\ }\bibfield
  {title} {\bibinfo {title} {{Search for high-mass dilepton resonances using
  139 fb$^{-1}$ of $pp$ collision data collected at $\sqrt{s}=$13 TeV with the
  ATLAS detector}},\ }\href {https://doi.org/10.1016/j.physletb.2019.07.016}
  {\bibfield  {journal} {\bibinfo  {journal} {Phys. Lett. B}\ }\textbf
  {\bibinfo {volume} {796}},\ \bibinfo {pages} {68} (\bibinfo {year} {2019})},\
  \Eprint {https://arxiv.org/abs/1903.06248} {arXiv:1903.06248 [hep-ex]}
  \BibitemShut {NoStop}%
\bibitem [{\citenamefont {Erler}\ \emph {et~al.}(2011)\citenamefont {Erler},
  \citenamefont {Langacker}, \citenamefont {Munir},\ and\ \citenamefont
  {Rojas}}]{Erler:2011ud}%
  \BibitemOpen
  \bibfield  {author} {\bibinfo {author} {\bibfnamefont {J.}~\bibnamefont
  {Erler}}, \bibinfo {author} {\bibfnamefont {P.}~\bibnamefont {Langacker}},
  \bibinfo {author} {\bibfnamefont {S.}~\bibnamefont {Munir}},\ and\ \bibinfo
  {author} {\bibfnamefont {E.}~\bibnamefont {Rojas}},\ }\bibfield  {title}
  {\bibinfo {title} {{Z' Bosons at Colliders: a Bayesian Viewpoint}},\ }\href
  {https://doi.org/10.1007/JHEP11(2011)076} {\bibfield  {journal} {\bibinfo
  {journal} {JHEP}\ }\textbf {\bibinfo {volume} {11}},\ \bibinfo {pages}
  {076}},\ \Eprint {https://arxiv.org/abs/1103.2659} {arXiv:1103.2659 [hep-ph]}
  \BibitemShut {NoStop}%
\bibitem [{\citenamefont {Salazar}\ \emph {et~al.}(2015)\citenamefont
  {Salazar}, \citenamefont {Benavides}, \citenamefont {Ponce},\ and\
  \citenamefont {Rojas}}]{Salazar:2015gxa}%
  \BibitemOpen
  \bibfield  {author} {\bibinfo {author} {\bibfnamefont {C.}~\bibnamefont
  {Salazar}}, \bibinfo {author} {\bibfnamefont {R.~H.}\ \bibnamefont
  {Benavides}}, \bibinfo {author} {\bibfnamefont {W.~A.}\ \bibnamefont
  {Ponce}},\ and\ \bibinfo {author} {\bibfnamefont {E.}~\bibnamefont {Rojas}},\
  }\bibfield  {title} {\bibinfo {title} {{LHC Constraints on 3-3-1 Models}},\
  }\href {https://doi.org/10.1007/JHEP07(2015)096} {\bibfield  {journal}
  {\bibinfo  {journal} {JHEP}\ }\textbf {\bibinfo {volume} {07}},\ \bibinfo
  {pages} {096}},\ \Eprint {https://arxiv.org/abs/1503.03519} {arXiv:1503.03519
  [hep-ph]} \BibitemShut {NoStop}%
\bibitem [{\citenamefont {Benavides}\ \emph {et~al.}(2018)\citenamefont
  {Benavides}, \citenamefont {Mu\~noz}, \citenamefont {Ponce}, \citenamefont
  {Rodr\'\i{}guez},\ and\ \citenamefont {Rojas}}]{Benavides:2018fzm}%
  \BibitemOpen
  \bibfield  {author} {\bibinfo {author} {\bibfnamefont {R.~H.}\ \bibnamefont
  {Benavides}}, \bibinfo {author} {\bibfnamefont {L.}~\bibnamefont {Mu\~noz}},
  \bibinfo {author} {\bibfnamefont {W.~A.}\ \bibnamefont {Ponce}}, \bibinfo
  {author} {\bibfnamefont {O.}~\bibnamefont {Rodr\'\i{}guez}},\ and\ \bibinfo
  {author} {\bibfnamefont {E.}~\bibnamefont {Rojas}},\ }\bibfield  {title}
  {\bibinfo {title} {{Electroweak couplings and LHC constraints on alternative
  Z' models in $E_6$}},\ }\href {https://doi.org/10.1142/S0217751X18502068}
  {\bibfield  {journal} {\bibinfo  {journal} {Int. J. Mod. Phys. A}\ }\textbf
  {\bibinfo {volume} {33}},\ \bibinfo {pages} {1850206} (\bibinfo {year}
  {2018})},\ \Eprint {https://arxiv.org/abs/1801.10595} {arXiv:1801.10595
  [hep-ph]} \BibitemShut {NoStop}%
\end{thebibliography}%
\end{document}